\documentclass[12pt]{article}
\usepackage{latexsym}
\usepackage{amstext}
\usepackage{amsmath}
\usepackage[normalem]{ulem}
\usepackage{subfigure}
\usepackage{epsfig, graphicx}
\usepackage{color}
\usepackage{amsfonts,bm,color}
\usepackage{verbatim}
\usepackage{float}
\usepackage{latexsym}

\newcommand{\ds}{\displaystyle}
\newcommand{\beq}{\begin{eqnarray}}
\newcommand{\eeq}{\end{eqnarray}}
\newcommand{\beqq}{\begin{eqnarray*}}
\newcommand{\eeqq}{\end{eqnarray*}}
\newcommand{\p}{\partial}

\newcommand{\eps}{\varepsilon}
\newcommand{\x}{\mbox{\boldmath$x$}}
\newcommand{\X}{\mbox{\boldmath$X$}}
\newcommand{\n}{\mbox{\boldmath$n$}}

\newcommand{\y}{\mbox{\boldmath$y$}}

\font\bb=msbm10 at 12pt
\def\rR{\hbox{\bb R}}

\begin{document}
\title{Poisson-Nernst-Planck equations in a ball}
\author{Z. Schuss\footnote{Department of Mathematics, Tel-Aviv University, Tel-Aviv 69978, Israel, {schuss@post.tau.ac.il}},
J. Cartailler and D. Holcman\footnote{Ecole Normale Sup\'erieure, 46 rue d'Ulm 75005 Paris, France, {david.holcman@ens.fr}.}}
\date{\today}

\maketitle
\begin{abstract}
The Poisson Nernst-Planck equations for charge concentration and electric potential in a ball is a model of electro-diffusion of ions in
the head of a neuronal dendritic spine. We study the relaxation and the steady state when an initial charge of ions is injected into the ball. The steady state equation is similar to the Liouville-Gelfand-Brat\'u-type equation with the difference that the boundary condition is Neumann, not Dirichlet and there a minus sign in the exponent of the exponential term. The entire boundary is impermeable to the ions and the electric field satisfies the compatibility condition of Poisson's equation. We construct a steady radial solution and find that the potential is maximal in the center and decreases toward the boundary. We study the limit of large charge in dimension 1,2 and 3. For the case of a small absorbing window in the sphere, we find the escape rate of an ion from the steady density.
\end{abstract}

\section{Introduction}
The non-linear system of Poisson-Nernst-Planck (PNP) equations has been widely used to study properties of the electric field in local nanodomains such as ionic channels \cite{nadlerschuss,Nadler,GHH1943,barcilon,baricloneisenberg}. It was also used to simulate the equilibration of ions between large reservoirs through narrow necks \cite{nadlerschuss,Nitzan}.  It is also possible to study the effect of interacting ions in ionic channels cite \cite{Taflia,Chen}.\\
We use here PNP to study the distribution of charges at the micrometer scale level. Indeed, the stationary PNP equations with Neumann and no-flux conditions on the boundary of a finite domain $\Omega$, respectively, describe the electrical potential and density of charge in $\Omega$. They can be reduced to a Liouville-Gelfand-Brat\'u-type equation for the electric potential with however two major differences: first, the boundary condition on $\p\Omega$ is Neumann but not Dirichlet and second, there is a minus sign in the exponent, which is normalized over the domain $\Omega$. We study here the solution of this equation in spherical symmetry in dimensions $\leq 3$ with respect to the (dimensionless) total charge $\lambda$.  We construct asymptotic approximations of the solutions for small and large $\lambda$. The one-dimensional case is solved explicitly and it is characterize by a log-singularity at the boundary that develop in the large $\lambda$ limit. The explicit solution in two-dimension has also a singularity on the boundary. We also obtain a similar asymptotic behavior in three-dimension, although the solution cannot be computed explicitly and we provide an asymptotic and numerical argument for large $\lambda$ showing again a log-singularity at the boundary. We study the voltage change from the center and how it develops a boundary layer for large $\lambda$. The voltage drop from the center to the boundary converges to a finite value as $\lambda$ increases to infinity.  We also apply the analysis of PNP to study idealized dendritic spine structure, which consists of a spherical dielectric membrane filled with ionic solution, connected to the dendrite by a cylindrical narrow neck.
The paper is organized as follow: in the first part we study asymptoticall and numericall PNP. In the second part, we estimate the current generate in an idealized spine (head connected by a cylinder). We derive the current generate by a spine. We show here how the head geometry controls the voltage, while the narrow neck radius control the current.

\section{PNP equations in a ball}\label{s:PNP-eq}
We consider the Poisson-Nernst-Planck system in a ball $\Omega$ of radius $R$, whose
dielectric boundary $\p\Omega$ is represented as the compatibility condition for
Poisson's equation and its impermeability to the passage of ions is represented as a
no-flux boundary condition for the Nernst-Planck equation. We assume that there are $N$ positive ions of valence $z$ in $\Omega$ and that there is an initial particle density $q(\x)$ in $\Omega$ such that
\begin{align}
\int\limits_\Omega q(\x)\,d\x=N.
\end{align}
The charge in $\Omega$ is
$$Q=zeN,$$
where $e$ is the electronic charge. The charge density $\rho(\x,t)$ is the solution of the Nernst-Planck equation
\begin{align}
D\left[\Delta \rho(\x,t) +\frac{ze}{kT} \nabla \left(\rho(\x,t) \nabla \phi(\x,t)\right)\right]=&\,
\frac{\p\rho(\x,t)}{\p t}\hspace{0.5em}\mbox{for}\ \x\in\Omega\label{NPE}\\
D\left[\frac{\p\rho(\x,t)}{\p n}+\frac{ze}{kT}\rho(\x,t)\frac{\p\phi(\x,t)}{\p n}\right]=&\,0\hspace{0.5em}\mbox{for}\ \x\in\p\Omega \label{noflux}\\
\rho(\x,0)=&\,q(\x)\hspace{0.5em}\mbox{for}\ \x\in\Omega,\label{IC}
\end{align}
where the electric potential in $\Omega$ is $\phi(\x,t)$ is the solution of
the Poisson equation
\begin{align}
\label{poisson} \Delta \phi(\x,t) =
-\frac{ze\rho(\x,t)}{\eps\eps_0}\hspace{0.5em}\mbox{for}\ \x\in\Omega
\end{align}
and the boundary condition
\begin{align}
\frac{\p\phi(\x,t)}{\p
n}=-\sigma(\x,t)\hspace{0.5em}\mbox{for}\ \x\in{\p\Omega},
\end{align}
where $\sigma(\x,t)$ is the surface charge density on the boundary $\p\Omega$. In the steady state
and in spherical symmetry
\begin{align}
\sigma(\x,t)=-\frac{Q}{4\pi R^2}.
\end{align}
\subsection{The steady-state solution}\label{ss:SSS}
In the steady state $\p\rho/\p t=0$ so (\ref{NPE}) gives the density
\begin{align}
\rho(\x)=N\frac{\exp\left\{-\ds\frac{ze\phi(\x)}{kT}\right\}}
{\ds\int_\Omega\exp\left\{-\ds\frac{ze\phi(\x)}{kT}\right\}\,d\x},\label{N}
\end{align}
hence \eqref{poisson} gives
\begin{align}
\Delta\phi(\x)=-\frac{zeN\exp\left\{-\ds\frac{ze\phi(\x)}{kT}\right\}}{\eps\eps_0{\ds\int_\Omega
\exp\left\{-\ds\frac{ze\phi(\x)}{kT}\right\}\,d\x}}.\label{Deltaphi}
\end{align}
In spherical symmetry in $\rR^d$ \eqref{Deltaphi} can be written in spherical coordinates as
\begin{align} \label{eqsymm1}
\phi''(r)+\frac{d-1}r\phi'(r)=
-\frac{zeN\exp\left\{-\ds\frac{ze\phi(r)}{kT}\right\}}{S_d\eps\eps_0{\ds\int_0^R
\exp\left\{-\ds\frac{ze\phi(r}{kT}\right\}\,r^{d-1}\,dr}}<0,
\end{align}
where $S_d$ is the surface area of the unit sphere in $\rR^d$. The boundary conditions are
\beq
\frac{\p\phi(0)}{\p r}=0,\quad\frac{\p\phi(R)}{\p r}=-\frac{Q}{S_d R^{d-1}}. \label{compatibility}
\eeq
The inequality in \eqref{eqsymm1} means that $\phi(r)$ has a maximum at the origin and decreases toward the boundary (see Fig.~\ref{f:LambdaS}A). We can normalize the radius by setting $r=Rx$ for $0<x<1$ and
\beq \label{conversion}
u(x)=\ds \frac{ze \phi(r)}{kT},\quad\lambda= \frac{(ze)^2N}{\eps\eps_0 kT},
\eeq
 to write \eqref{eqsymm1} as
\begin{align}
u''(x)+\frac{d-1}x u'(x)=&\, -\frac{\lambda \exp \left\{-\ds u(x) \right\}}{{S_dR^{d-2}\ds\int_0^1
\exp\left\{-\ds u(x))\right\}\, x^{d-1}dx}}\label{eqsymm}\\
u(0)=&\,0,\quad u'(0)=0.\nonumber
\end{align}
Incorporating the denominator of the RHS of \eqref{eqsymm} into the parameter $\lambda$ by setting
\begin{align}
\lambda=\mu S_dR^{d-2}\int_0^1\exp\{-u(x)\}\,x^{d-1}\,dx,\label{lambdamu}
\end{align}
we can write the initial value problem \eqref{eqsymm} as
\begin{align}
u''(x)+\frac{d-1}x u'(x)=& -\mu \exp \left\{-\ds u(x) \right\}\label{eqmu}\\
u(0)=&\,u'(0)=0.\nonumber
\end{align}
First, we show that solutions exist in dimensions $1\leq d\leq3$ only for $\mu$ in the range $0\leq\mu<\mu^*$ for some positive
$\mu^*$.
\subsubsection*{Solution in dimension one}
We solved directly equation \ref{eqmu} in dimension 1 (see appendix \ref{ss:SSS}) and we obtain  (see eq. \ref{u1lambdax})that
\beq
u^{1D}_{\lambda}(x)=\ln    \cos^{2}\left( \sqrt {\frac{\lambda}{2I_{\lambda}}}x\right), \label{u1lambdaxtext}
\eeq
where $I_{\lambda}$ is solution of the implicit equation
\beq
I_{\lambda}= {\frac{2}{\lambda}} \tan^2 \sqrt {\frac{\lambda}{2I_{\lambda}}}.
\eeq
The graph of $u^{1D}_{\lambda}(x)$ is shown  Fig. \ref{f:1D_2D_3D}A, while the one for  $\frac{\lambda}{I_{\lambda}}$ versus $\lambda$ is shown in Fig. \ref{f:1D_2D_3D}B. We have $0<\mu(\lambda)=\frac{\lambda}{I_{\lambda}}\leq \frac{\pi^2}{2}$ and $\lim_{\lambda \rightarrow \infty} \mu(\lambda) =\frac{\pi^2}{2}$. The solution exists $u^{1D}_{\lambda}$ for all $\lambda>0$ and a log-singularity develops at the boundary $x=1$ when $\lambda \rightarrow \infty$.
\subsubsection*{Solution in dimension two}
In dimension 2, we obtain the solution in  (appendix \ref{ss:SSS2})
\beq
u_{\lambda}^{2D}(x)=\log {(1-\frac{\lambda}{8I_{\lambda}} x^2)^2}.
\eeq
where
\beqq
I_{\lambda}&=&\pi+\frac18 \lambda\\
\mu(\lambda)&=& \frac{\lambda}{I_{\lambda}}\\
\lim_{\lambda\rightarrow \infty}\mu(\lambda) &=&8.
\eeqq
The graph of $u_{\lambda}(x)$ is shown on Fig. \ref{f:1D_2D_3D}C, while the one for $\frac{\lambda}{I_{\lambda}}$ is on Fig. \ref{f:1D_2D_3D}D.
$u_{\lambda}(x)=\log (1-\frac{\lambda}{\lambda+8\pi} x^2)^2$, develop a log-singularity as $\lambda \rightarrow \infty$.
\subsubsection*{Analysis in dimension three}
The solution of the initial value problem \eqref{eqsymm} in dimension $d=3$ can be directly computed. We show now that the solution exits for all $\lambda$, while there is a critical value $\mu^*$, above which, there is no regular solution. Contrary to dimensions  one and two, the value of $\mu^*$ can only be estimated numerically. We first show using a phase-space analysis that the solution of equation \ref{eqmu} is unique when it exists. However it is not possible to use the phase-space to study the singularity of the equation. To study the asymptotic explosion of the equation, we use an asymptotic argument. Finally, we will study the solution numerically.

\begin{figure}[http!]
\center
\includegraphics[scale=1]{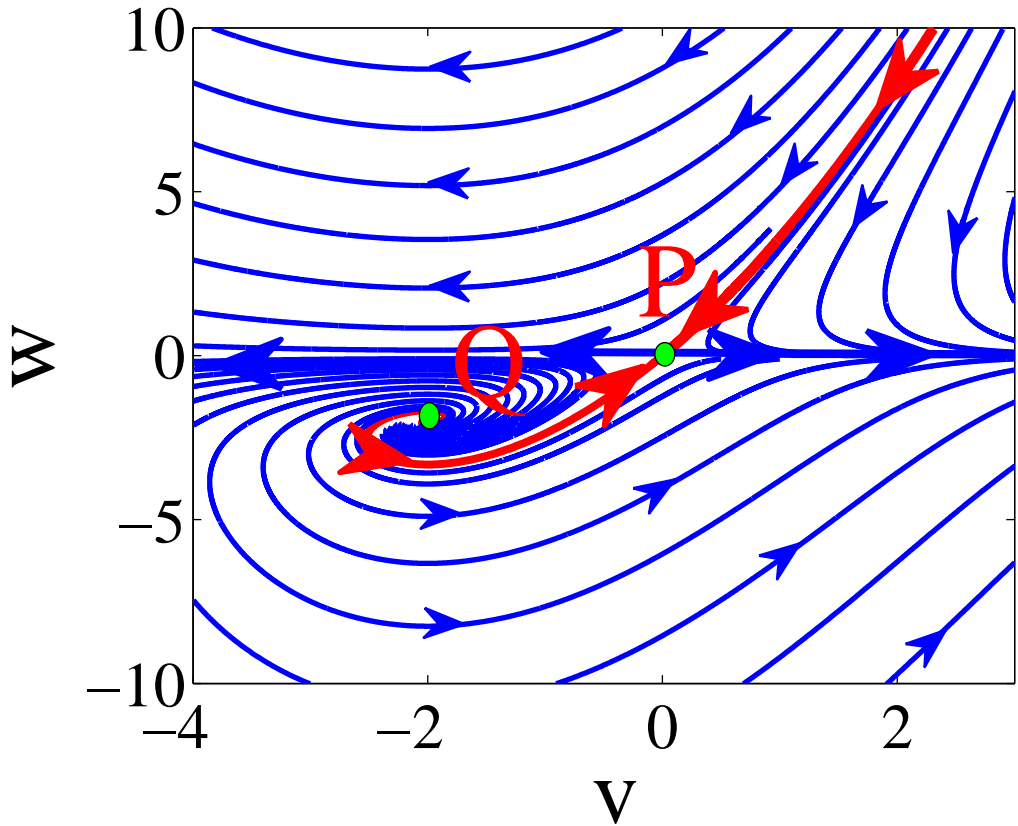}
\caption{\small {\bf Phase-space solution of \eqref{phs1}}. The separatrix is shown in red, while the other trajectories are in blue.}
\label{f:phaseP}
\end{figure}

Next, we show that the problem \eqref{eqsymm} has a  unique regular solution for all $\lambda\geq0$, when the solution is finite. The proof of uniqueness of the solution follows the phase-space analysis of \eqref{eqmu}. Indeed,  using the change of variables
\begin{align}
s=&-\log r,\hspace{0.5em}u(r)=U(s),\hspace{0.5em} v(s)=\frac{dU(s)}{ds},\hspace{0.5em} w= \mu e^{-2s}e^{-U(s)}\nonumber\\
w'(s)=&-2w(s)-U'(s)w(s)=w(s)[-2-v(s)],\label{cov}
\end{align}
which gives
\begin{align}
v'(s)=v(s)-w(s),\quad w'(s)=-w(s)[2+v(s)],\label{phs1}
\end{align}
and can be written as
\begin{align}
\frac{dw}{dv}=\frac{-w(2+v)}{v-w}.\label{de1}
\end{align}
The phase space of \eqref{phs1} contains exactly two critical points. The origin $\bf 0$ is a saddle point and its stable manifold has the tangent ${\bf T}$ of equation $w=3v$. The point $P_a= (-2,-2)$ is an unstable node. The initial conditions $u(0)=u'(0)=0$ for the solution of \eqref{eqmu} impose $\lim_{s\to\infty}U(s)=u(0)=0$ and $\lim_{s\to\infty}U'(s)=-\lim_{r\to0}ru'(r)=v(0)=0$, hence the constraints
\beq
\lim_{s\rightarrow\infty}v(s)=0,\quad\lim_{s\rightarrow\infty}w(s)=\lim_{s\to\infty}\mu e^{-2s}e^{-U(s)}=0.\label{constraints}
\eeq
Thus the trajectory of the solution of \eqref{eqmu} in the first quadrant, which satisfies the constraints \eqref{constraints}, has to be on the separatrix that converges to the saddle point. Choosing any value $U(0)$ gives $\mu e^{-U(0)}$ the value of $v(0)=U'(0)$ has to be chosen on the separatrix.  Therefore starting in the first quadrant, a trajectory of \eqref{phs1} converges to the saddle point if and only if it starts on the separatrix with the tangent ${\bf T}$. The stable branch at the saddle point tends to infinity  as $s$ decreases toward $0$. Indeed, the local expansion of \eqref{de1} near the saddle point is
\beq\label{local}
w(v)=3v+\frac{3}{5}v^2-\frac{3}{175}v^3+\ldots,
\eeq
which gives the phase portrait (Fig. \ref{f:phaseP}). Finally, along the separatrix $w'(v)>0$, except at the origin, showing that for an initial  $v(0)$, there is a unique solution. However, it is not possible from the phase-space to study singular solution. Indeed, as we shall see, when there is a singularity,  because the it occur precisely at the initial value and thus the Cauchy problem cannot even start.
We conclude the problem \eqref{eqsymm} has a finite solution and the phase diagram plotted in Fig. \ref{f:phaseP} ensures that for any initial condition $(v(0),w(0))$ (when it is finite) on the separatrix in the first quadrant, there is a unique solution to \eqref{phs1} that satisfies \eqref{constraints}.

\begin{figure}[http!]
\center
\includegraphics[scale=0.5]{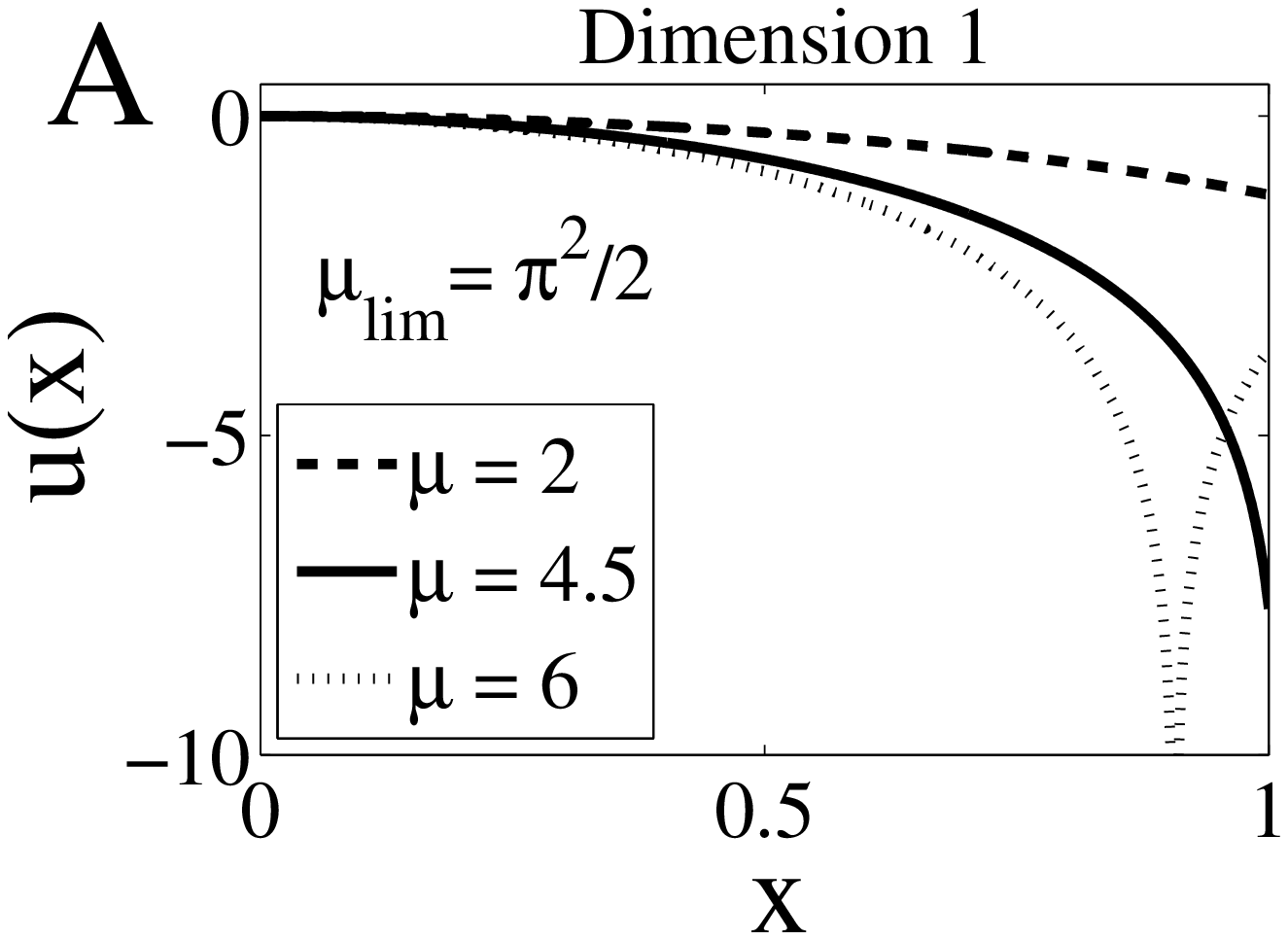}
\includegraphics[scale=0.5]{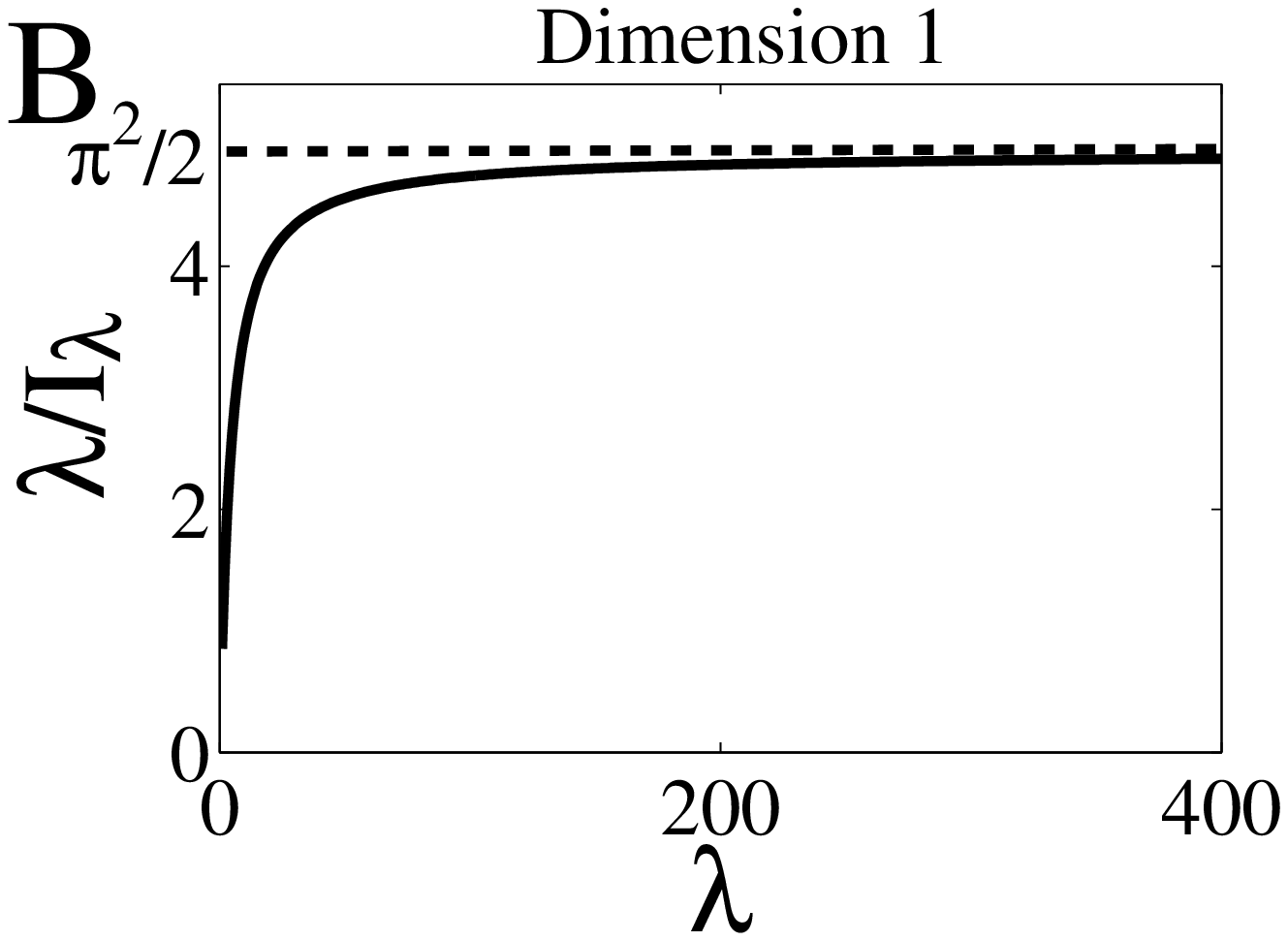}
\includegraphics[scale=0.5]{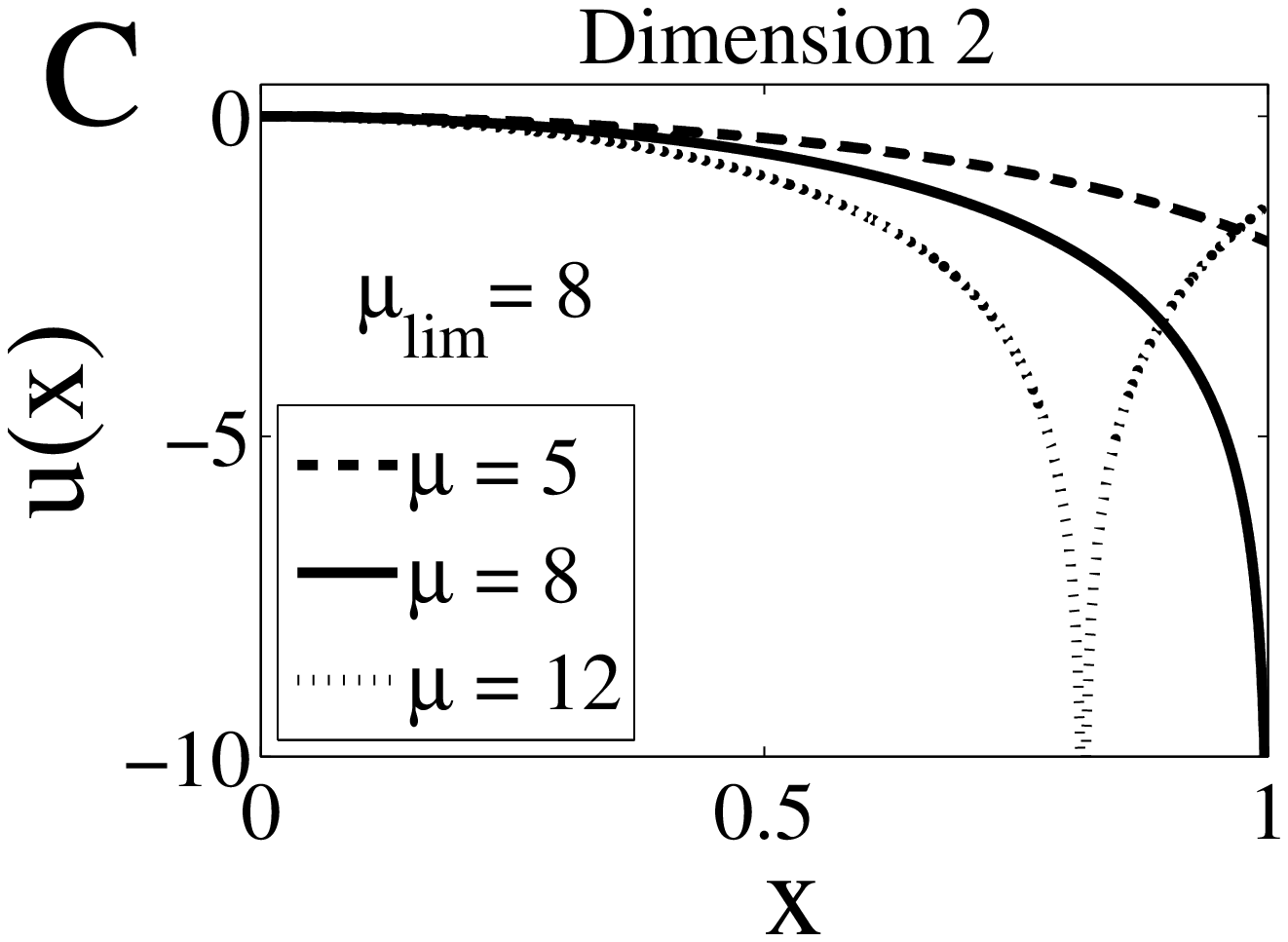}
\includegraphics[scale=0.5]{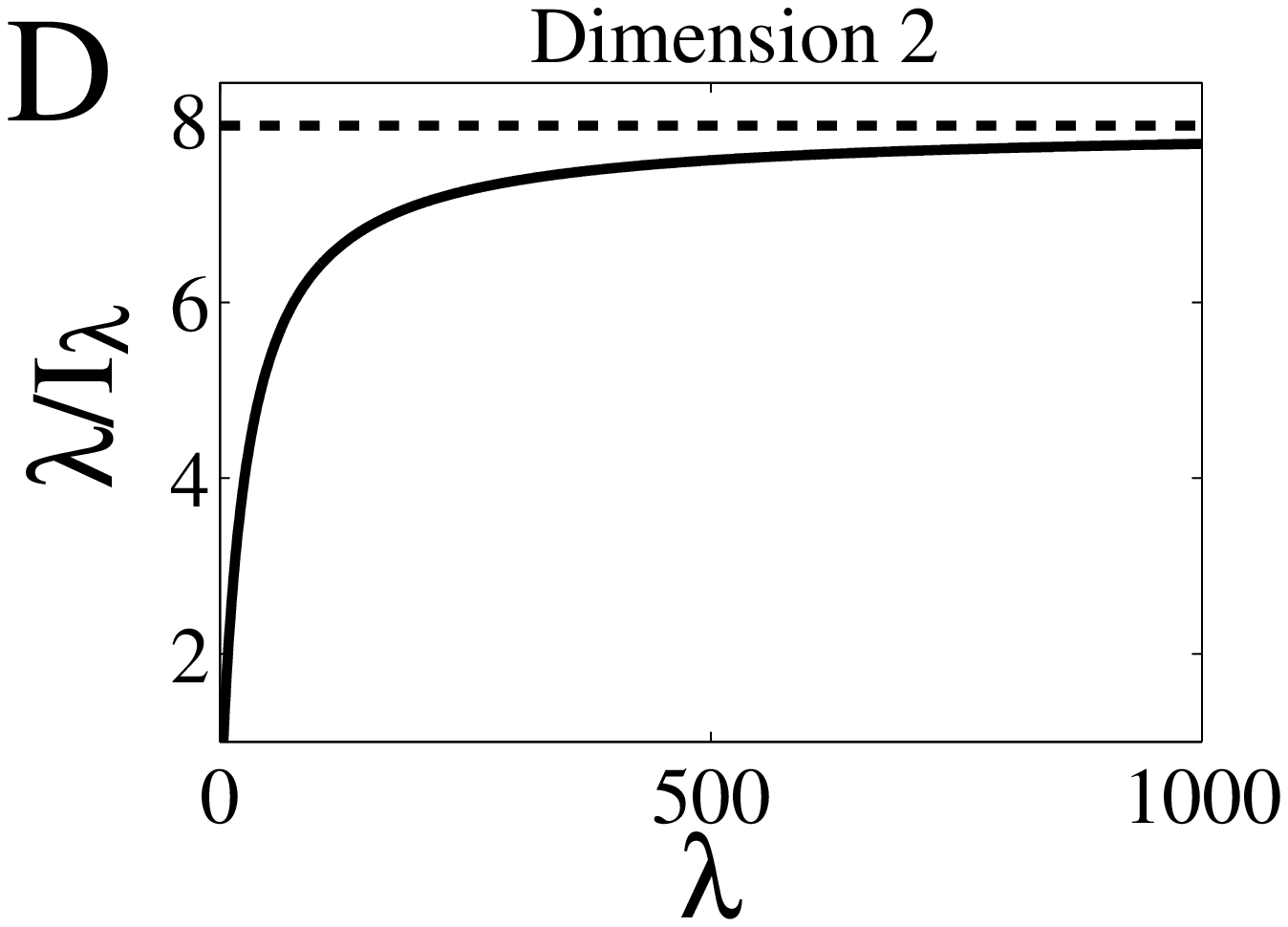}
\includegraphics[scale=0.5]{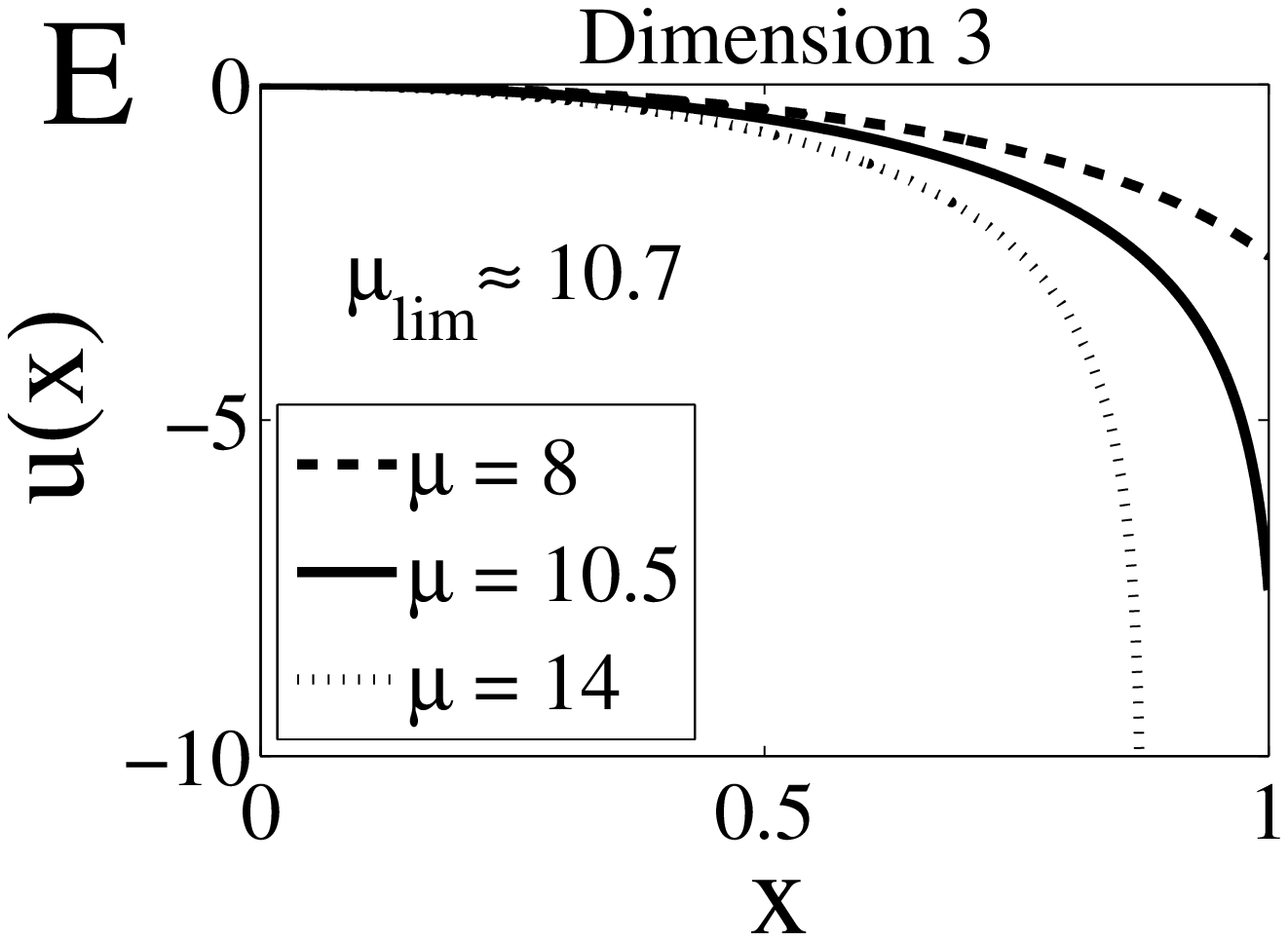}
\includegraphics[scale=0.5]{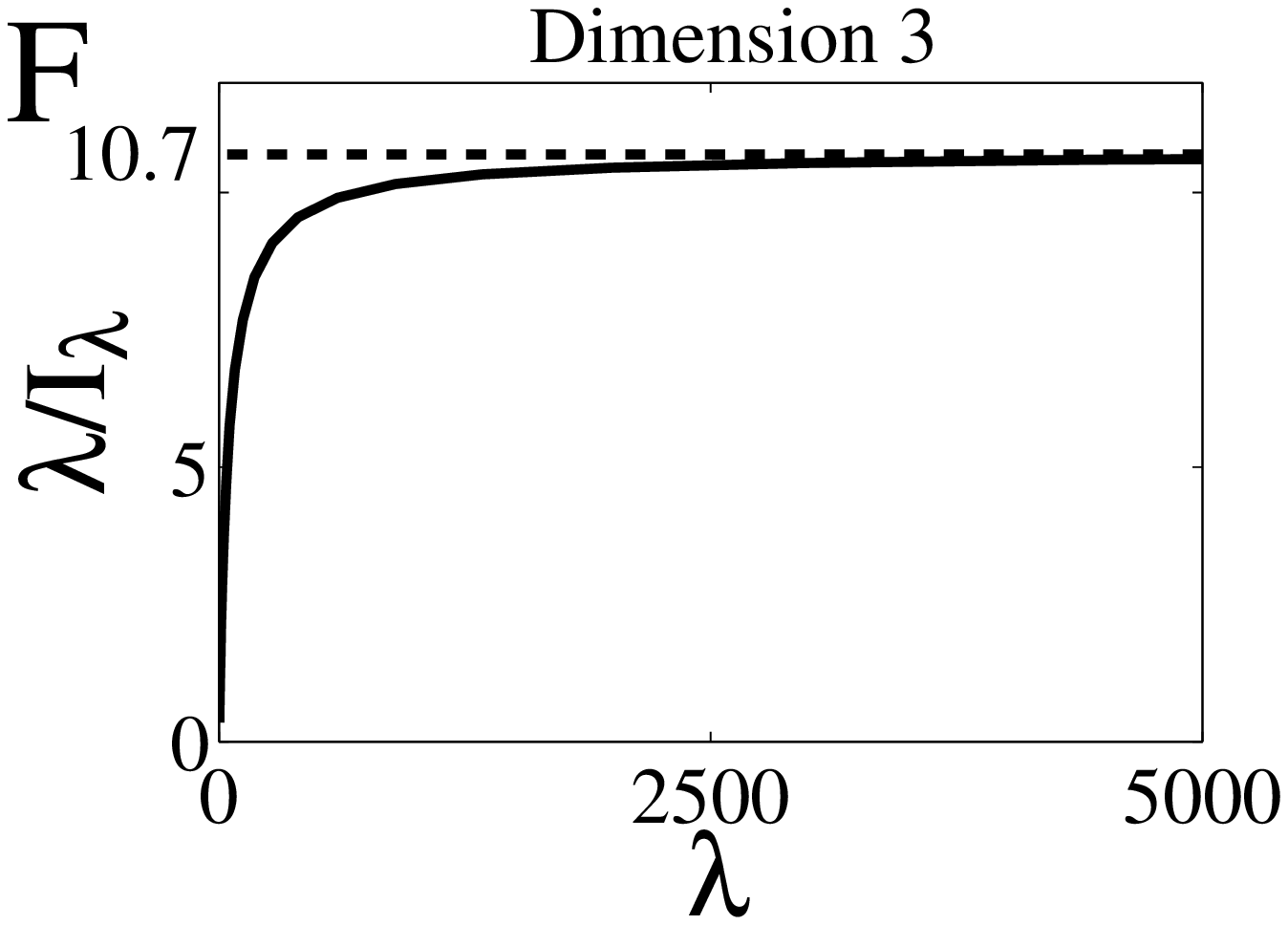}
\caption{\small {\bf Numerical solutions $u(x)$ of the initial value problem \eqref{eqsymm}}. (A),(C), and (E) correspond to different profile values of $\lambda$ in dimensions 1,2, and 3, respectively. The dotted curves are solutions that blow-up for $x<1$. (B),(D), and (F) are plots of the ratio $\ds \frac{\lambda}{I_{\lambda}}$ in dimensions 1,2 and 3, respectively.}
\label{f:1D_2D_3D}
\end{figure}

\begin{figure}[http!]
\center
\includegraphics[scale=0.5]{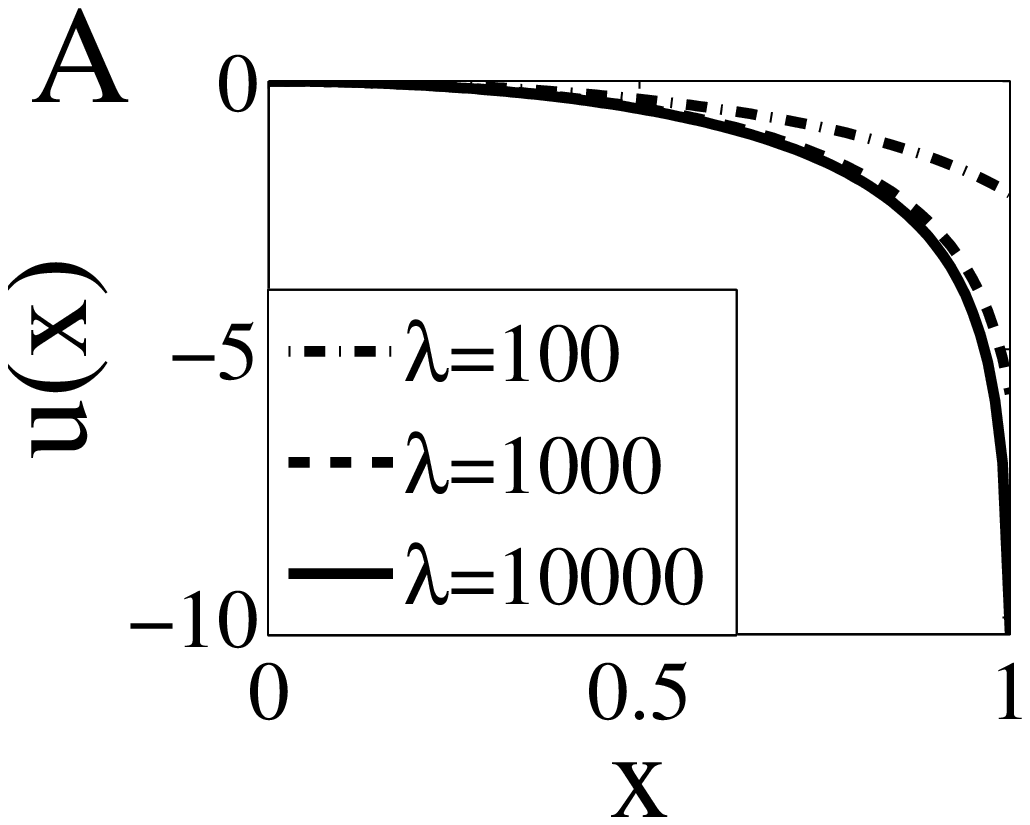}
\includegraphics[scale=0.5]{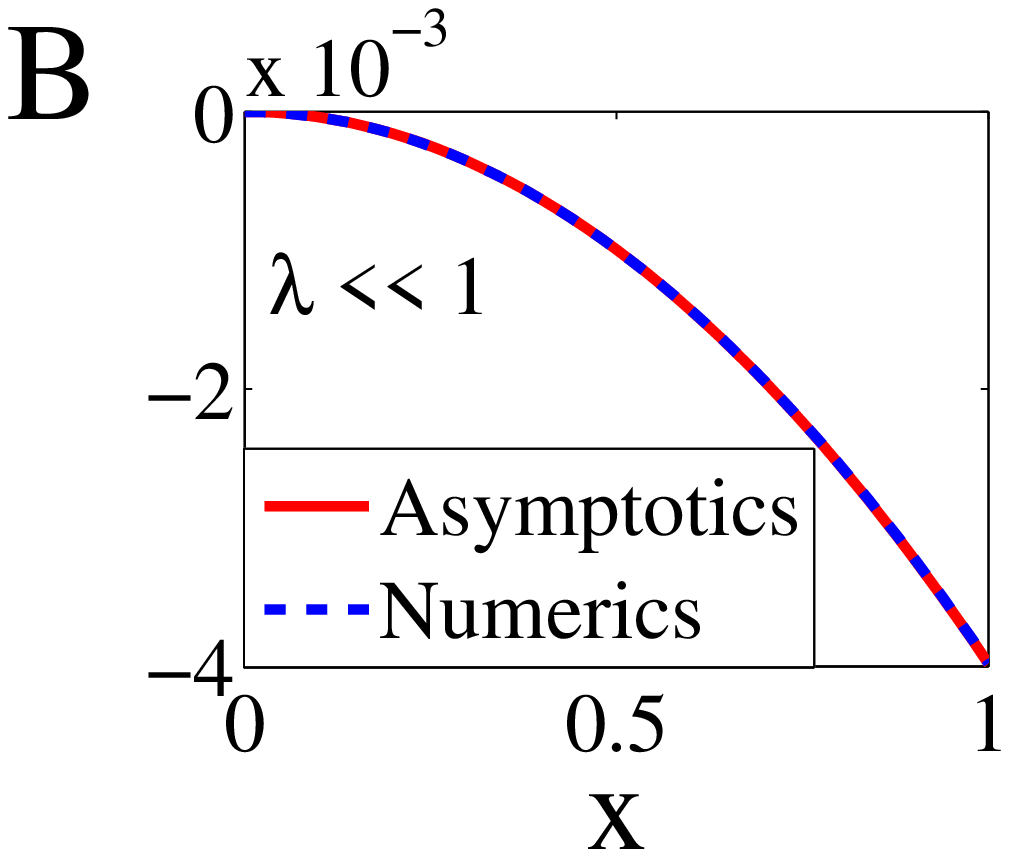}
\includegraphics[scale=0.5]{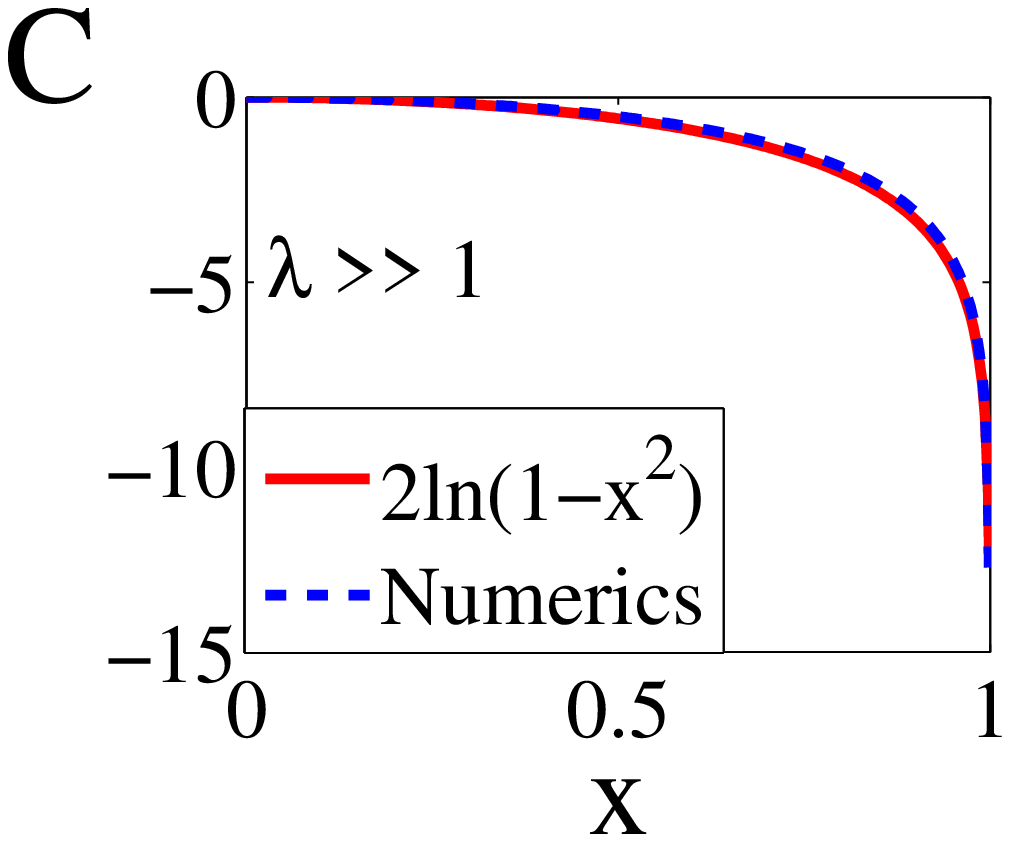}
\caption{{\bf Asymptotics behavior of the solution $u(x)$.} {(\bf A)} change in the profile $u(x)$ for 3 values of the parameter $\lambda=10^2, 10^3,10^4.$  {(\bf B,C)} We present two regimes: for $\lambda=0.1\ll 1$, we have  $u(x)=-\lambda\frac{x^2}{8\pi}+O(\lambda^2)$ (see eq. \ref{ux}) and $\lambda \gg 1$ where $u(x)\approx 2\ln(1-x^2)$. The analytical approximations (red) are compared with the numerical solutions (see appendix). }
\label{f:LambdaS}
\end{figure}

A numerical solution of \eqref{eqsymm} gives the graph Fig.~\ref{f:LambdaS}E, that is the solution $u(x)$ of \eqref{f:LambdaS} for $\mu\leq \mu^*=11.2$ . The graph in dashed line ($\mu^*=14$) blows up before reaching $x=1$, while the dash (small point) graph is finite throughout the interval. To estimate an upper bound for $\mu^*$, we note that whenever the solution exists for some $\mu$ near $\mu^*$, its asymptotic behavior for $x$ close to $1$ shows that $u''(1)\gg u'(1)$ (see the blue graph in Fig. \ref{f:LambdaS}). Indeed, to show that under the assumption $u''(1)\gg u'(1)$ the latter inequality is self-consistent, we note that near $x=1$ the solution of \eqref{eqsymm} can be
approximated  by the solution of the simpler problem
\begin{align}
\tilde u''(x)= -\mu \exp \left\{-\ds \tilde u(x) \right\},\label{eqmuv}
\end{align}
given by
\begin{align}
\tilde u(x)\sim\log\cos^2\left(\sqrt{\frac{\mu}{2}}\,x\right).\label{vrm}
\end{align}
Thus {$\tilde u(x)$ } is finite in the interval as long as
\begin{align}
\mu<\frac{\pi^2}{2}=4.934802202=\mu^*
\end{align}
and
\begin{align}
\frac{\tilde u'(x)}{\tilde u''(x)}\leq\frac{|\sqrt{\mu}-\sqrt{\mu^*}|}{\sqrt{\mu^*}}\ll1.
\end{align}
We conclude at this stage that for fixed values of $\mu$ below and above, where above the latter they blow-up inside the interval $0<x<1$ (frames A,C,E of Fig. \ref{f:1D_2D_3D}). When $\mu$ varies with $\lambda$ according to \eqref{lambdamu}, the solutions exist for all values of $\lambda$ (frames B,D,F of Fig. \ref{f:1D_2D_3D}). Figure \ref{f:1D_2D_3D}A-C-E shows that potential drop between the center and the surface of the sphere as a function of $\lambda$ for $1\leq d\leq3$.\\
In figure \ref{f:LambdaS}, we compare the three dimensional solution obtained numerically with the asymptotic expansions in two regimes. We present in appendix \ref{smalllabmda} for $\lambda \ll 1$, the expansion $u(x)=-\lambda\frac{x^2}{8\pi}+O(\lambda^2)$ (see eq. \ref{ux}). In contrast, for  $\lambda \gg 1$, we mention above that the approximation $u(x)\approx 2\ln(1-x^2)$, which was relevant near $x=1$ can be used in the entire domain $[0,1]$. The analytical approximations (red) are compared with the numerical solutions (see appendix \ref{numsol}).
\subsubsection*{The potential differences }\label{ss:profile}
\begin{figure}[http!]
\center
\includegraphics[scale=0.5]{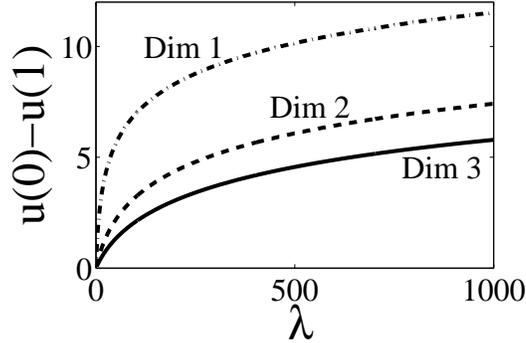}
\caption{Asymptotics of $u_{\lambda}(1) -u_{\lambda}(0)$ for dimensions 1,2 and 3.}
\label{f:drop}
\end{figure}

The difference $u(0)-u(1)$ as we shall see in the next section has a physical meaning, as it represents the difference of potential between the center and the periphery of a sphere.  We have in dimension 1,
\beq
\mid u_{\lambda}(1) -u_{\lambda}(0)\mid&=&\ln    \cos^{2}\left( \sqrt {\frac{\lambda}{2I_{\lambda}}}\right),
\eeq
where $\frac{\lambda}{2I_{\lambda}} \rightarrow \frac{\pi^2}{4}$ as $\lambda \rightarrow\infty$.
in dimension 2,
\beq
\mid u_{\lambda}(1) -u_{\lambda}(0)\mid&=&2\log(\frac{8\pi}{\lambda+8\pi} ),
\eeq
in dimension 3, for $\lambda\gg1$,
\beq
\mid u_{\lambda}(1) -u_{\lambda}(0)\mid&=&2\log\ln(1-f(\lambda)),
\eeq
where the function $f$ is increasing and $f(\lambda) \rightarrow 1$ as $\lambda \rightarrow\infty$. The different curves for dimension 1,2 and 3 are shown in fig. \ref{f:drop}. In all cases, the large $\lambda$ asymptotic is dominated by the log-behavior.
\subsection{Physical implication for the distribution of voltage and charge in a dielectric ball}
\begin{figure}[http!]
\center
\includegraphics[scale=0.5]{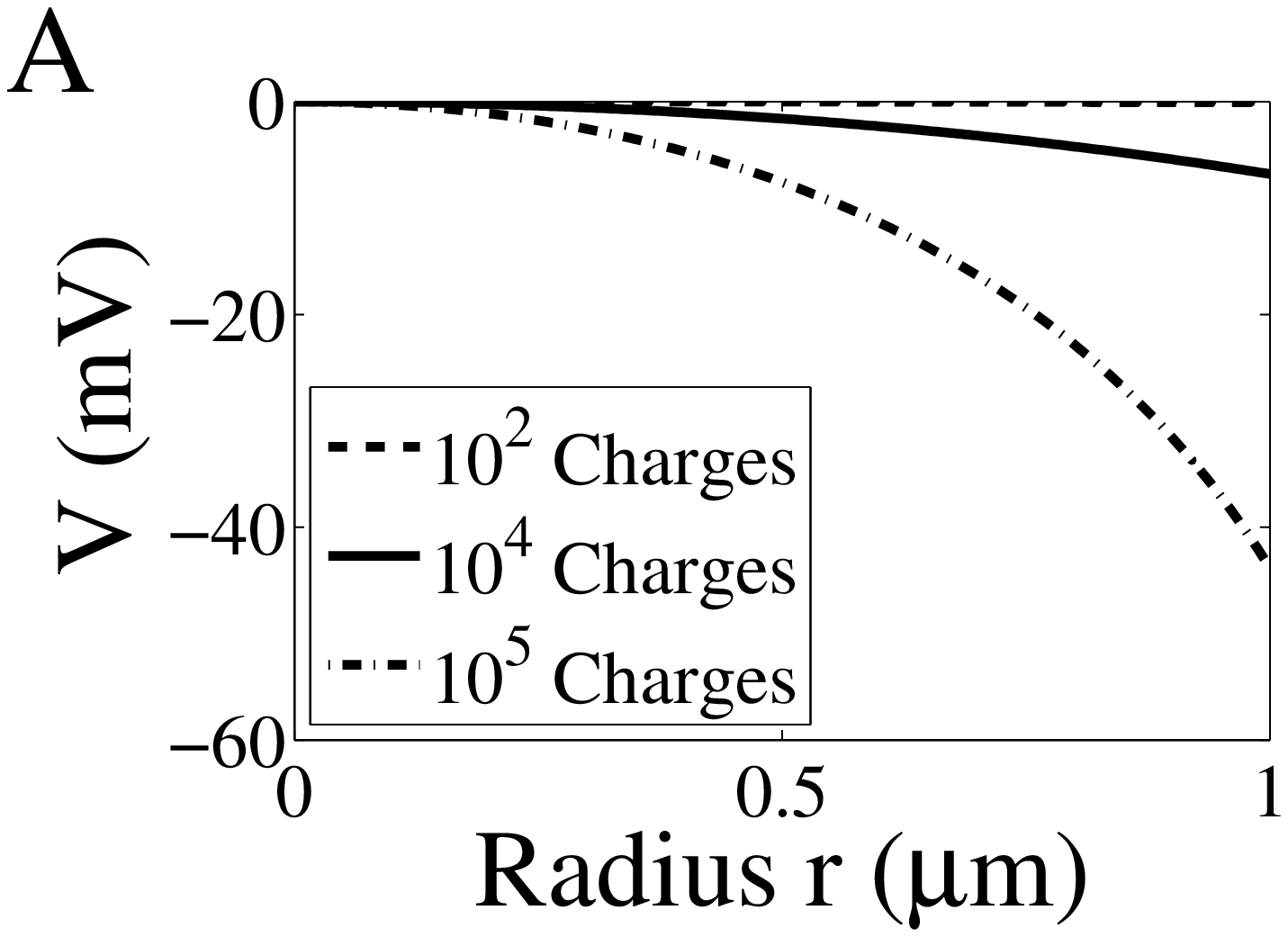}
\includegraphics[scale=0.5]{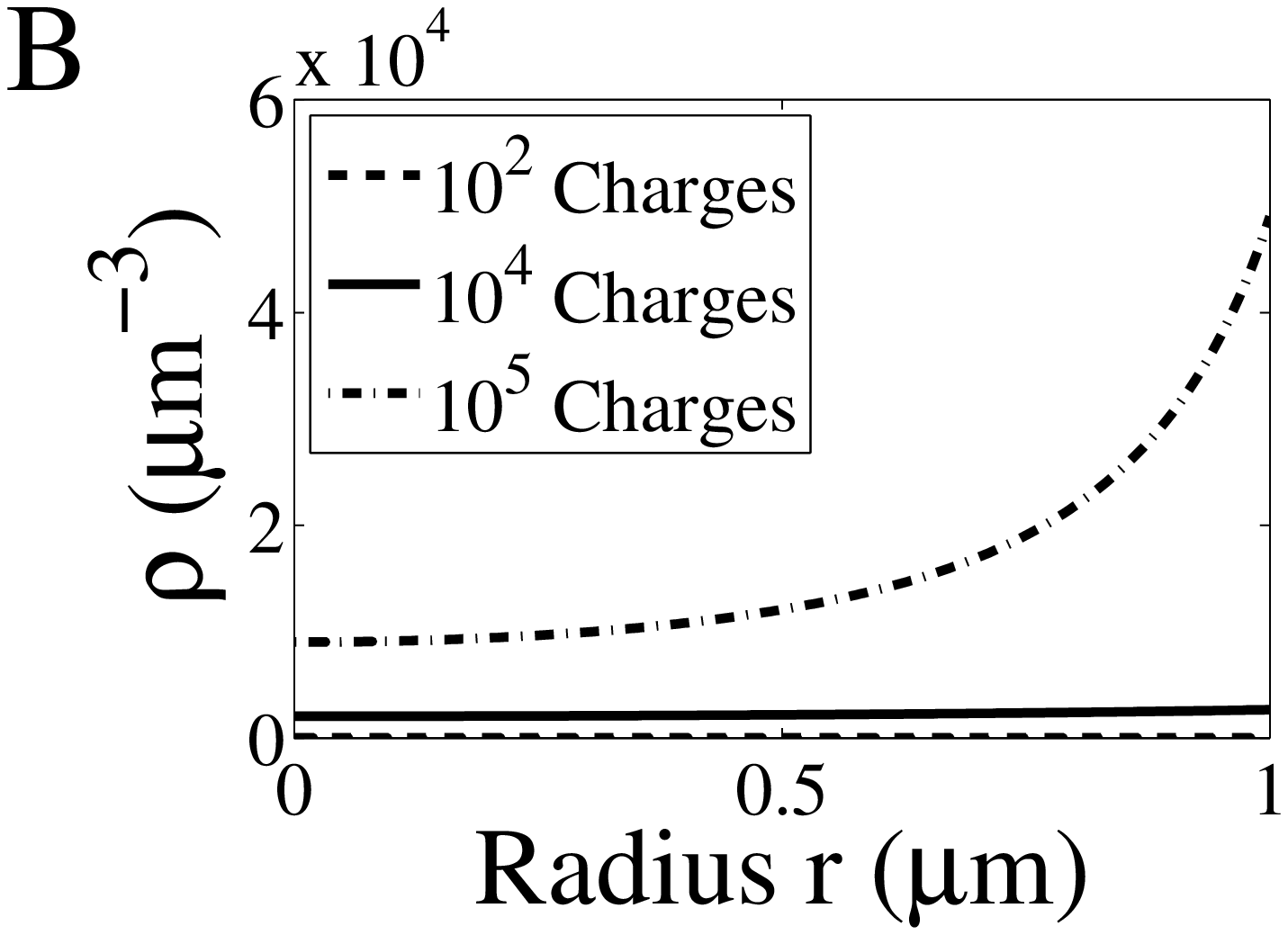}
\includegraphics[scale=0.5]{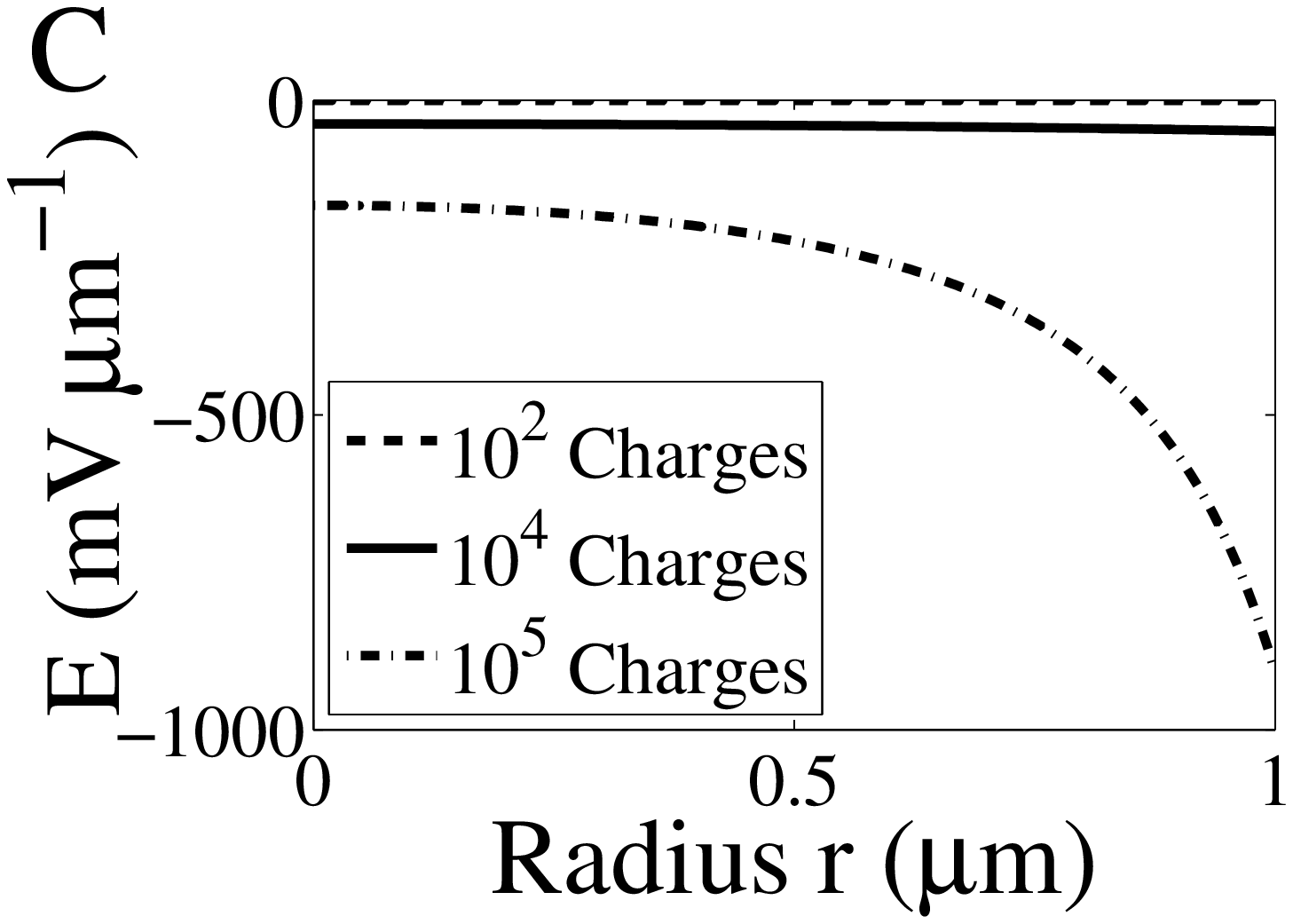}
\includegraphics[scale=0.5]{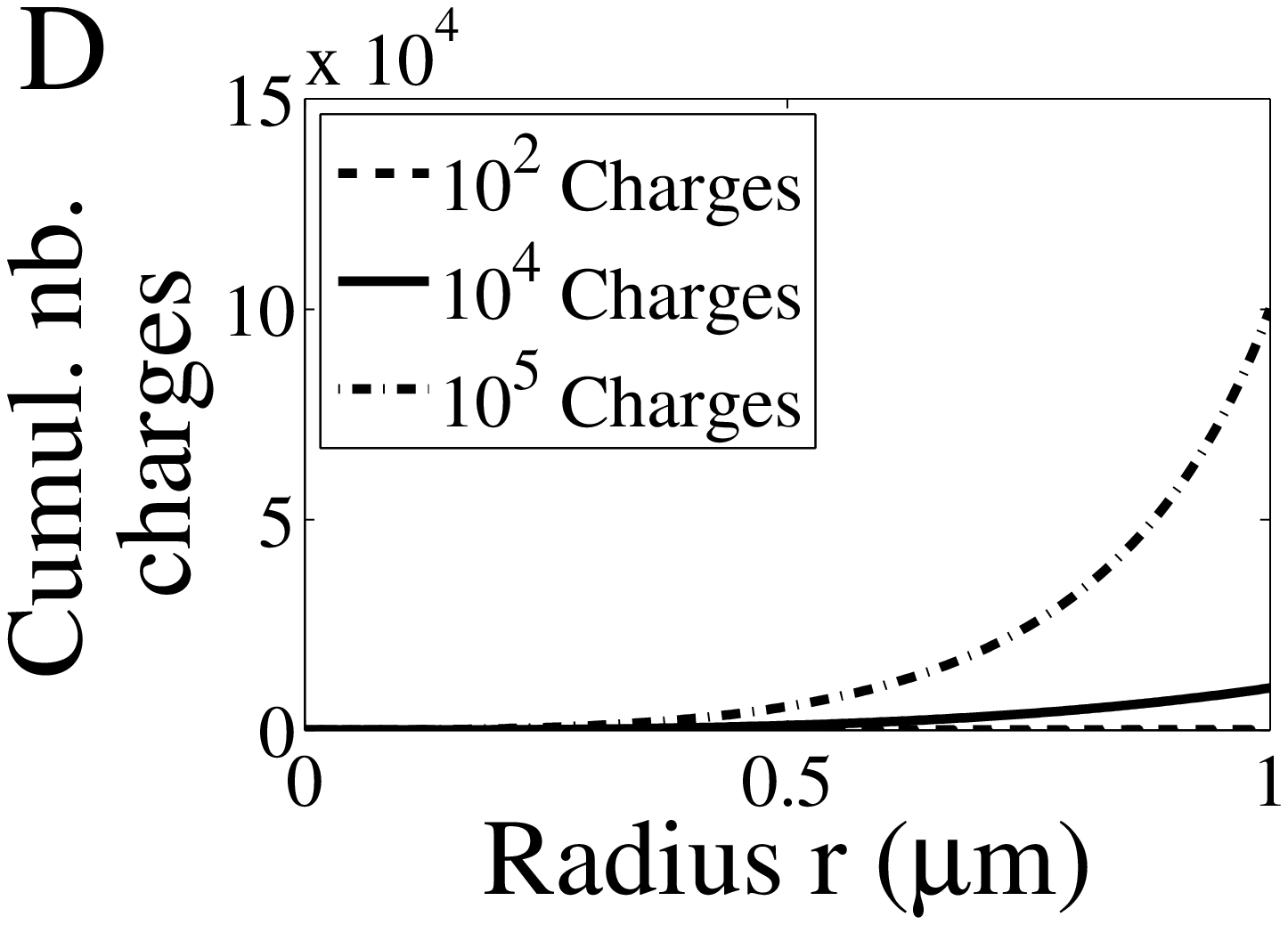}
\caption{{\bf Distribution of {(\bf A)} the potential, {(\bf B)} charge and the field {(\bf C)} and cumulative density of charges {(\bf D)} inside a dielectric ball.}}
\label{f:Physical}
\end{figure}

The distribution of voltage and charge in a dielectric ball can be estimated from the results of the previous results by using the dimensional relation \ref{conversion} in a ball of radius $R=1 \mu m $. We plotted in Fig. \ref{f:Physical}A the voltage drop for $N=10^2,10^3 10^4$ charges. Already for 1000 charges, there is a difference between the center and the surface of a ball of few milli-Volt. This effect could be tested for in the head of dendritic spines. Moreover, the density is charge is concentrated at the periphery (Fig. \ref{f:Physical}B), leading also to a large field close to the boundary (Fig. \ref{f:Physical}C). Consequently most od the charge are accumulated at the boundary, as revealed by the plot of the cumulative density of charges
\beq
Q(r)&=&N \frac{\ds\int_0^r  \exp\left\{-\frac{ze\phi(r)}{kT}\right\}\,r^2\,dr}{\ds\int_0^R \exp\left\{-\frac{ze\phi(r)}{kT}\right\}\,r^2 \,dr}.
\eeq
We conclude that when the total number of charges is fixed sufficiently high, the charge accumulate at the surface. The field is only significant close to the surface and thus can trap a Brownian charge in such region, while outside a small boundary layer of the boundary, the field is almost zero and charge particle experience no drift. This effect is discussed in the section.
\subsection{Scaling laws for the maximum number of charges}
Although we found previously that for a fixed radius, the difference of potential $V(0)-V(1)$ is bounded as a function of the total number of charge, we shall now show that the maximal number of charges increases linearly with the radius of the ball. Indeed, introducing the dimensionless radial variable $\zeta=r/R$ and $u_\lambda(r)=U_{\lambda/R}(\zeta)$, equation (\ref{eqsymm1}) becomes
\begin{align} \label{eqsymm2}
U_{\lambda/R}''(\zeta)+\frac2\zeta \,U_{\lambda/R}'(\zeta)= -\frac{\lambda \exp \left\{-\ds U_{\lambda/R}(\zeta)\right\}}{{4\pi R\ds\int_0^1
\exp\left\{-\ds U_{\lambda/R}(\zeta)\right\}\, \zeta^2\,d\zeta}},
\end{align}
with the initial conditions $U_{\lambda/R}(0)=U_{\lambda/R}'(0)=0$. Now, we solve the initial value problem
\begin{align}
V_\mu''(\zeta)+\frac2\zeta \,V_\mu'(\zeta)=&\, -\mu \exp \left\{-\ds V_\mu(\zeta)\right\},\quad V_\mu(0)=V_\mu'(0)=0\nonumber\\
&\label{eqsymm3}\\
W'_\mu(\zeta)=&\,\zeta^2\exp \left\{-\ds V_\mu(\zeta)\right\},\quad W_\mu(0)=0\nonumber
\end{align}
and note that
\begin{align}
u_\lambda(r)=V_\mu\left(\frac{r}{R}\right),\quad\lambda=4\pi\mu RW(1).
\end{align}
Thus the number of charges $Q$ in a ball or radius $R$ create the same distribution as a charge $Q/R$ in a ball of radius one, which can be summarized as
\beq
Q(R)=RQ(1).
\eeq

\section{Ionic flux in a small absorbing window in a highly charged sphere}
We now discuss various consequences of distributing charges close to the boundary, in the large charge regime. The first consequence is on the MFPT $\bar\tau(\x)$ from $\x\in\Omega$, which is the solution of the Pontryagin-Andronov-Vitt (PAV) boundary value problem \cite{DSP}
\begin{align}
D\left[\Delta \bar\tau(\x) -\frac{ze}{kT}\nabla \bar\tau(\x)\cdot \nabla \phi(\x)\right]=&-1
\hspace{0.5em}\mbox{for}\ \x\in\Omega\label{PAV}\\
\frac{\p \bar\tau(\x)}{\p n}+\frac{ze}{kT}\bar\tau(\x)\frac{\p\phi(\x)}{\p n}=&0
\hspace{0.5em}\mbox{for}\ \x\in\Omega_r\label{reflection}\\
\bar\tau(\x)=&0\hspace{0.5em}\mbox{for}\ \x\in\Omega_a.\label{absorption}
\end{align}
We consider the case of a large field $-\nabla\phi(\x)\gg1$ near the boundary $|\x|=1$. The profile of $\phi(\x)$ was studied in section \ref{ss:profile} (see Figures \ref{f:Physical}). To study the solution of the PAV problem \eqref{PAV}-\eqref{absorption}, we map the neighborhood of  $\p\Omega_a$ smoothly into the upper half plane with coordinates $\X=(x,y,z)$, where $z=0$ is the image of the boundary, $\tilde \tau(\X)=\bar\tau(\x)$, and outside a boundary layer near $\p\Omega_a$
$$V=\frac{\p\phi(\x)}{\p n}\Big{|}_{|\x|=1}=const,\quad \Phi(x,y)=\phi(\x)\Big{|}_{|\x|=1}=const,$$
so that $\nabla_{x,y}\Phi(x,y)=0.$
The PAV system \eqref{PAV}-\eqref{absorption} is converted to
\begin{align}
\tilde u\tau_{zz}(\X) -\frac{ze}{kT} V\tilde \tau_z(\X)+\Delta_{x,y} \tilde \tau(\X)=-\frac{1}{D},
\end{align}
A regular expansion of $\tilde u(\X)$ for large $V$ gives that to leading order $\tilde \tau(\X)$ is a function of $(x,y)$ and setting $T(x,y)=\tilde \tau(x,y,0)$, we find that
\begin{align}
\Delta_{x,y} T(x,y)=-\frac{1}{D}.\label{Txy}
\end{align}
Thus the MFPT from $\x\in\Omega$ to $\p\Omega_a$ is the sum of the MFPT from $\x$ to $\p\Omega$ and the MFPT form $\p\Omega$ to $\p\Omega_a$ on the surface $\p\Omega$. The MFPT to $\p\Omega$ is negligible relative to that to $\p\Omega_a$. This approximation means that to reach $\p\Omega_a$ in a highly charged ball a charge is first transported by the field to the reflecting part $\p\Omega_r$ of the sphere with overwhelming probability and then it finds $\p\Omega_a$ by surface diffusion.
\subsection{The current through a small absorbing window in a highly charged sphere}
A second consequence of the charge distributions is the control of spine current, independently of the voltage. Indeed, the solution $T(x,y)$ of \eqref{Txy} is the MFPT of Brownian motion on a sphere of radius $R$  to an absorbing circle centered at the north-south axis near the south pole, with small radius $a = R \sin\frac{\delta}{2}$. It is given by \cite{NarrowEscape3}
\beq
T(x,y) = \frac{2R^2}{D}\log\frac{\sin \frac{\theta}{2}}{\sin\frac{\delta}{2}},
\eeq
where $D$ is the diffusion coefficient, $\theta$ is the angle between $\x$ and the north pole. Thus
\beq
\bar\tau(\x)=T(x,y).
\eeq
The MFPT, averaged over the sphere with respect to a uniform distribution of $\x$ is given by
\begin{align}
\bar\tau=2R^2\left(\log\frac{1}{\delta}+O(1)\right)\hspace{0.5em}\mbox{for}\ \delta\ll1.
\end{align}
The MFPT for $N$ independent charges is
\begin{align}
\bar\tau_N=\frac{2R^2}{N}\left(\log\frac{1}{\delta}+O(1)\right)\hspace{0.5em}\mbox{for}\ \delta\ll1.
\end{align}
It follows that the current through the small window is given by
\begin{align}
J=\frac{ze}{\bar\tau_N}=\frac{Q D}{2R^2\left(\log\ds\frac{R}{a}+O(1)\right)}\hspace{0.5em}\mbox{for}\ a\ll R.
\end{align}
We conclude that once a current enters into a dielectrics ball such as a spine head, the excess of charges $Q$ is first pushed toward the boundary and before moving by Brownian motion to the spine neck. This result shows that the current in a spine head is governed by the spine geometry and a key parameter is the radius $a$ of the neck. When there is a conservation of charge principle (no leak), the current through the dendritic shaft is the same as the one exiting the spine. In that conditions, the spine neck length do not affect or modulate the current.
\section{The current in a spine neck under voltage-clamp condition}
Determining the voltage drop between the membrane of the spine head and the dendrite when a current is flowing from the head to the dendrite remains challenging because the cable theory cannot be applied in a system that cannot be approximated by a cable. The general scheme for modeling the electro-diffusion in the spine is the PNP model in the head and a one-dimensional conduction of ions in the neck. The neck is considered a classical ionic conductor. Thus the steady-state PNP equations have to be solved in the sphere with boundary conditions implied by the compatibility condition and the flux through the neck is determined by the mean first passage time (MFPT) of ions from the head to the neck, as discussed above. In the case of high charge $Q$ the potential turns out to be practically flat throughout the ball with a sharp boundary layer with negative slope at the boundary. Thus charge diffuses and is pushed strongly toward the membrane so ionic motion is practically confined to motion on the surface. Due to spherical symmetry, the potential is constant on the boundary so ionic motion is free Brownian motion on a sphere. At high charge ions interact through the ambient potential that is determined from Poisson's equation in the ball. Therefore they can be assumed independent free Brownian particles. The MFPT $\bar\tau$ of an ion to the small opening of the neck is determined from the two-dimensional NET theory(see previous section). Because the flux carried by a single ion is $q/\bar\tau$, where $q$ is the ionic charge, the number of ions in the spine head is $N=Q/q$ and the MFPT $\bar\tau_N$ of any of the $N$ ions is given by
$$\bar\tau_N=\frac{\bar\tau}{N}.$$
Thus the current through the neck is
\beq
I=\frac{Q}{\bar\tau}
\eeq
and due to charge conservation, it is independent of the length of the neck. If we consider the neck to be a parallel-plate capacitor carrying a steady current $I$, then the voltage drop across the neck is simply $V=RI$, where $R$ is the resistance of the neck, given by
\beq
R=\frac{k_BTL^2}{q^2nD},
\eeq
where $k_B$ is Boltzmann's constant, $T$ is absolute temperature, $L$ is the length of the conductor, $n$ is the number of ions in the neck, $q$ is the charge of an ion, and $D$ is the diffusion coefficient of the solution in the neck \cite{DSP}. This model is valid as along as the voltage is maintained in the spine head.
\section{Discussion, application and conclusion}
We have studied here the solution of the PNP equations in a ball. We estimated the voltage in dendritic spines when the voltage in the spine head is maintained. A certain fraction of spines receive synaptic connections, essential for neuronal communication. Although their functions are still unclear, there are involved in regulating synaptic transmission and plasticity \cite{Yuste,segal,Svoboda,Sabatini}. Interestingly, most of the excitatory connections occurs not on the dendrite but rather on spines and the reason is still not clear. The spine shape is quite intriguing, made of a head connected to the dendritic shaft by a cylinder. We found here that this geometry play a key role: the spine head geometry determines the drop of potential, while the current is defined by the diffusion on the surface and the mean time to find the entrance of the neck in a two dimensional Brownian motion (see Narrow escape time \cite{PNAS}\cite{JPA2014}). In the neck, under a voltage clamp condition, when a constant voltage difference between the head and the neck is imposed, the voltage-current relation follow a resistance law. Thus the spine geometry defines both the capacitance and resistance in geometrical terms, a vision that complement previous classical studies \cite{Svoboda,Koch,Segev}.

Finally, computing in the transient regime, the change in voltage drop between the spine head and the dendritic shaft, requires computing the time dependent PNP equations. Another open question is to study the influence of the spine head geometry on the distribution of charge. Computing the distribution of charges and the associated field in non-convex geometry is certainly the most challenging.
\section{Appendix}\label{numsol}
In this appendix, we first solve analytically the Liouville equation \ref{eqsymm} in dimensions one and two and in the second part, we describe the numerical methods to compute the solution in dimension 3.
\subsection{Solution of the minus sign Liouville-Bratu-Gelfand equation in a unitary segment}\label{ss:SSS}
Liouville equation in the interval  [0 1] is
\beq \label{DeltaphiLGBB}
-u''(r)&=&\lambda \frac{e^{-u}}{\int_{0}^1e^{-u}\,dr}
\eeq
with initial  conditions
\beq
u(0)=0 \hbox{ and }u'(0)=0.
\eeq
This is the classical Cauchy problem. After a direct integration we get with the initial conditions
\beq \label{DeltaphiLGBB}
u'^2(x)&=& \frac{2\lambda}{I_{\lambda}}(e^{-u(x)}-1),
\eeq
where
\beq
I_{\lambda}=\int_{0}^1e^{-u_{\lambda}(x)}dx.
\eeq
A second integration gives
\beq
u_{\lambda}(x)=\ln    \cos^{2} \sqrt {\frac{\lambda}{2I_{\lambda}}}x .\label{u1lambdax}
\eeq
Now we self-consistently calculate
\beq
I_{\lambda}=\int_{0}^1e^{-u_{\lambda}(x)}dx=\int_{0}^1 \frac{dx}{\cos^2\sqrt {\frac{\lambda}{2I_{\lambda}}}}= \frac{1}{\sqrt {\frac{\lambda}{2I_{\lambda}}}} \tan \sqrt {\frac{\lambda}{2I_{\lambda}}} .
\eeq
Thus $I_{\lambda}>0$ is the solution of the implicit equation
\beq\label{implicit}
I_{\lambda}= {\frac{2}{\lambda}} \tan^2 \sqrt {\frac{\lambda}{2I_{\lambda}}}.
\eeq
The graph of $\frac{\lambda}{I_{\lambda}}$ versus $\lambda$ is shown in Figure \ref{f:1D_2D_3D}. We have $\lim_{\lambda\to\infty}\frac{\lambda}{I_{\lambda}}=\frac{\pi^2}{2},$
and specifically, $y_\lambda=\sqrt{\frac{\lambda}{2I_{\lambda}}}=\frac{\pi}{2}-\frac{\pi^2}{\lambda^2}+O(\frac1{\lambda^2})$. The solution \eqref{u1lambdax} is shown in Fig. \ref{f:1D_2D_3D} and is regular in the entire interval $0<x<1$ for all values of $\lambda$. The drop between the extreme points of the interval is
\beq
u_{\lambda}(1)-u_{\lambda}(0)= \ln \cos^{2} \sqrt {\frac{\lambda}{2I_{\lambda}}}
\eeq
and becomes infinite as the total charge increases indefinitely.
\subsection{Liouville equation in dimension 2}\label{ss:SSS2}
The dimension 2 case can be transformed into the one dimensional case \cite{Jacobsen} using the change of variable
\beqq
r&=&e^{-t}\\
\tilde u(t)&=&u(r)-2t.
\eeqq
Equation \ref{eqsymm} reduces to
\beq \label{DeltaphiLGBBp}
-\tilde  u_{tt}&=& \frac{\lambda}{I_{\lambda}}e^{- \tilde u(t)+2t}
\eeq
where $I_{\lambda}= 2\pi \int_{0}^1 e^{-u(r)}rdr$ and $w(t)= \tilde u(t)+2t$ satisfies
\beq \label{DeltaphiLGBBp}
-w_{tt}&=&\lambda \frac{e^{-w(t)}}{I_{\lambda}}
\eeq
The initial condition are now transform to asymptotic conditions at infinity:
\beq
\lim_{t\rightarrow \infty}\left( w(t)-2t\right)&=&0\\
\lim_{t\rightarrow \infty}\left( \dot{w}(t)-2\right)e^{t}&=&0
\eeq
A first integration gives
\beq
\frac{\dot{w}^2}{2}=\lambda \frac{e^{-w(t)}}{I_{\lambda}}+2.
\eeq
The solution is
\beq
{w}(t)=-\log(\frac{8}{(\lambda e^{2C+2t}-1)^2}) -2C-2t,
\eeq
where  $C$ is a constant. Finally, we obtain that
\beq
u_{\lambda}(r)=\log {(1-\frac{\lambda}{8I_{\lambda}} r^2)^2}.
\eeq
To close the equation, we shall now compute the integral
\beq
I_{\lambda}=\int_{0}^1e^{-u_{\lambda}(r)}2\pi rdr=\int_{0}^1 \frac{1}{(1-\frac{\lambda}{8I_{\lambda}} r^2)^2} 2\pi rdr=\frac{8\pi}{8-\lambda/I_{\lambda}}
\eeq
and
\beq
I_{\lambda}&=&\pi+\frac18 \lambda\\
\lim_{\lambda\rightarrow \infty}\frac{\lambda}{I_{\lambda}}&=&8.
\eeq
The curve $\frac{\lambda}{I_{\lambda}}$ is shown on Fig. \ref{f:1D_2D_3D} and $\mid u_{\lambda}(1) -u_{\lambda}(0)\mid$ in Fig. \ref{f:drop}. Finally,
\beq
u_{\lambda}(r)&=&\log (1-\frac{\lambda}{\lambda+8\pi} r^2)^2 \\
\mid u_{\lambda}(1) -u_{\lambda}(0)\mid&=&2\log(1-\frac{\lambda}{\lambda+8\pi} ).
\eeq
We conclude that $u_{\lambda}(r)$ decreases smoothly and in the limit  $\lambda\rightarrow \infty$, the solution blow-up over the entire boundary.
\subsection{Regular expansion of solution \ref{eqmu} for small $\lambda$}\label{smalllabmda}
We shall now study the small asymptotic expansion of solution \ref{eqmu} for small $\lambda$. Using a regular expansion,
\beq
u(\x)=u_0(\x)+u_1(\x)\lambda +u_2(\x)\lambda^2 +o(\lambda^2),
\eeq
we obtain using eq. \ref{eqmu} that $u_0(\x)=0$ and $u_1$ is solution of
\beq
-\Delta u_1 &=&\frac{1}{\ds |\Omega|} \hbox{ on } \Omega \\
\frac{\p u_1}{\p \n}&=&-\frac{1}{\ds |\p \Omega|}  \hbox{ on } \p\Omega.
\eeq
For $R=1$,
\beq \label{ux}
u_1(r)&=&-\frac{r^2}{8\pi}.
\eeq
with $u_1(0)=0$. We conclude that $u_1(r)\leq 0$, Thus,
\beq
u(r)=-\frac{r^2}{8\pi}\lambda +O(\lambda^2).
\eeq
The second order term $u_2$ is solution of
\beq
-\Delta u_2 &=&-\frac{u_1}{\ds |\Omega|} \hbox{ on } \Omega,
\eeq
with $u_2(0)=0$ and $u_2'(0)=0$.
For $R=1$,
\beq \label{ux}
u_2(r)&=&-\frac{3r^4}{640\pi^2}.
\eeq
Thus,
\beq
u(r)=-\frac{r^2}{8\pi}\lambda -\frac{3r^4}{640\pi^2}\lambda^2+O(\lambda^3).
\eeq

\end{document}